# On-Chip Erbium-Doped Tantalum Oxide Microring Hybrid Cavity Single-Mode Laser

**Like Shui[1], Ruixuan yi[2], Chenyang Zhao[3], Jinlong Lu[1], Xiaotong Zhang[1], Jiaqiao Zhang[1], and Xuetao Gan[1,2*]**

[1] *School of integrated circuits and Microelectronics, Northwestern Polytechnical University, Xi'an 710129, China*
[2] *Key Laboratory of Light Field Manipulation and Information Acquisition, Ministry of Industry and Information Technology, and Shaanxi Key Laboratory of Optical Information Technology, School of Physical Science and Technology, Northwestern Polytechnical University, Xi'an 710129, China*
[3] *Analytical & Testing Center, Northwestern Polytechnical University, Shaanxi 710072, China*
*\*xuetaogan@nwpu.edu.cn*

**Abstract:** We demonstrate a high-performance, single-mode Er:Ta$_2$O$_5$ microring laser monolithically integrated on a silicon platform via a customized Damascene process. The Er:Ta$_2$O$_5$ gain medium exhibits a low propagation loss of 0.73 dB/cm and a high intrinsic Q-factor of 5.03×10$^5$. By utilizing a hybrid cavity—consisting of a microring coupled to a U-shaped waveguide at two symmetric points—we exploit the Vernier effect to achieve robust longitudinal mode selection. Under a non-resonant 1480 nm pumping scheme, the laser yields a side-mode suppression ratio (SMSR) of 53.3 dB and a narrow linewidth of 9.5 pm. A slope efficiency of 2.76 % is achieved—the highest reported to date for Er:Ta$_2$O$_5$ lasers—with a lasing threshold of ~ 3.3 mW. Furthermore, stable single-mode tuning is demonstrated across a temperature range of 18–68 °C, consistently aligning with theoretical transfer matrix models. This work provides a scalable pathway for high-efficiency, tunable on-chip light sources, bridging the gap for monolithic active-passive integration on the tantalum oxide photonic platform.

## 1. Introduction

In the field of integrated photonics, high-performance on-chip light sources serve as core components supporting critical applications such as optical communications[1-4], photonic sensing[5-8], and microwave photonics[9-12]. Their miniaturization, low loss, and high energy conversion efficiency have become central objectives of current research and development efforts[13-15]. Silicon photonic platforms, leveraging the advantages of mature manufacturing processes and high integration density, have long been important carriers for on-chip light sources[16]. Meanwhile, oxide materials doped with rare-earth ions, owing to their excellent optical gain characteristics, stand as ideal gain media for constructing silicon-based integrated lasers[17-20]. Among these materials, erbium-doped tantalum oxide (Er:Ta$_2$O$_5$) not only exhibits good compatibility with silicon-based processes but also demonstrates significant photoluminescent properties in the 1500–1577 nm communication band[21-23]. Additionally, it possesses advantages such as a high refractive index and low optical loss, thereby providing a reliable material foundation for achieving high-performance on-chip laser output[24].

Despite these advantages, the performance of current Er:Ta$_2$O$_5$ lasers is hindered by critical bottlenecks, particularly regarding energy conversion efficiency. Existing architectures, such as Fabry-Pérot (FP) cavities formed by depositing reflective coatings on straight waveguide facets, have demonstrated limited slope efficiencies (typically 0.3)[25]. Alternatively, microdisk resonators suffer from even lower conversion efficiencies due to sub-optimal optical field confinement and inefficient energy extraction[26]. Beyond efficiency, structural design and fabrication limitations remain prevalent. Traditional straight-cavity designs struggle to achieve high-purity single-mode operation with narrow linewidths and high side-mode suppression ratios (SMSR). Furthermore, conventional fabrication processes often encounter issues such as incomplete waveguide filling and high sidewall roughness, which induce significant scattering losses and degrade overall device performance[27].

To address the aforementioned issues, the academic community has attempted to improve device performance by optimizing resonant cavity structures or modifying fabrication processes[28-30]; however, existing schemes still have limitations. Some studies have improved mode selection effects by adjusting cavity parameter, but failed to achieve in-depth adaptation to the characteristics of Er:Ta$_2$O$_5$ materials. Other studies have optimized individual processes such as magnetron sputtering and annealing to reduce material losses, but have not fully leveraged the advantages of silicon-based large-scale manufacturing processes (e.g., the Damascene process), resulting in limited potential in terms of device integration density and customized design[31-33].

In this work, we demonstrate a fully packaged on-chip microring laser monolithically integrated on an erbium-doped tantalum oxide (Er:Ta$_2$O$_5$) platform compatible with silicon photonics. A hybrid resonant cavity consisting of a microring resonator and a U-like waveguide is constructed on a silicon wafer with a 10 μm thermally oxidized SiO$_2$ layer[34], resulting in a compact footprint of only 6.2 × 2.3 mm for the fabricated Er:Ta$_2$O$_5$ laser. This Er:Ta$_2$O$_5$ microring laser enables stable single-mode laser emission around 1556 nm when pumped by a 1480-nm laser diode, delivering a maximum on-chip output power of 72.14 μW from a single output port. More importantly, to the best of our knowledge, this work achieves a slope efficiency of 2.76%, which is significantly higher than the previously reported slope efficiencies of Er:Ta$_2$O$_5$ lasers[25,26]. This result highlights the potential to develop high-performance on-chip Er:Ta$_2$O$_5$ lasers with enhanced energy conversion efficiency through rational structural design and process optimization.

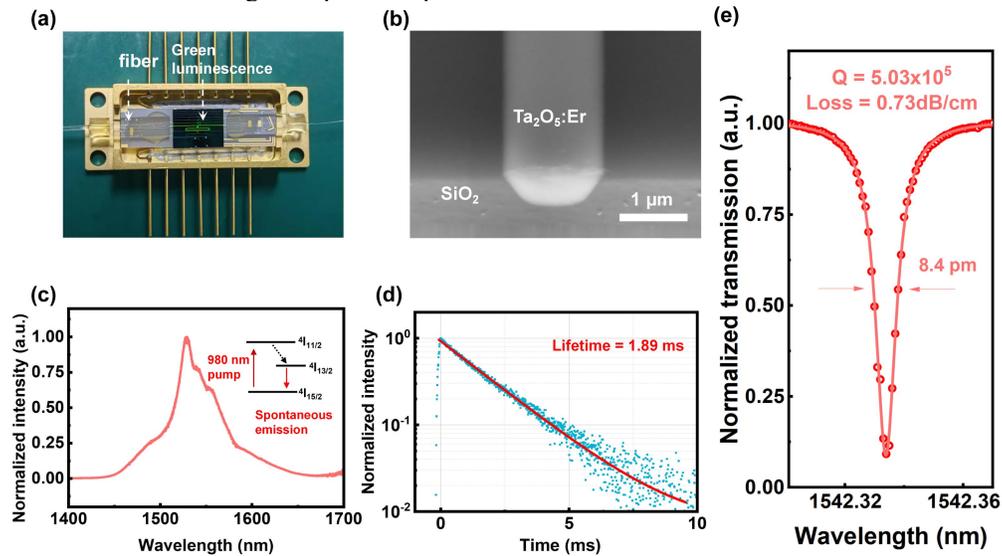

**Fig. 1 a** Optical image of the fully packaged fully packaged integrated erbium-doped tantalum oxide laser. **b** SEM image of the cross-section of the 1 μm erbium-doped tantalum oxide waveguide. **c** Photoluminescence Diagram of Erbium-Doped Tantalum Oxide Material Pumped by 980nm. **d** Fluorescence Lifetime Diagram Excited by 980nm. **e** Transmission Spectra Diagram of the All-pass Ring Resonator Fitted by Loren.

Fig. 1**a** illustrates the fully packaged erbium-doped tantalum oxide (Er:Ta$_2$O$_5$) laser. The device was fabricated on a silicon (Si) substrate featuring a 10 μm thermally grown SiO$_2$ buffer layer. The fabrication employed a Damascene process tailored for the Er:Ta$_2$O$_5$ photonic platform, comprising electron beam lithography (EBL), inductively coupled plasma

(ICP) etching, and a subsequent thermal reflow step to minimize sidewall roughness. The Er:$Ta_2O_5$ gain medium was deposited via magnetron sputtering, followed by chemical mechanical polishing (CMP) for planarization and high-temperature annealing for $Er^{3+}$ activation.As shown in the cross-sectional view (Fig. 1**b**), the magnetron sputtering was performed at 200 °C with an RF power of 100 W using a $Ta_2O_5$:$Er_2O_3$ (99:1 wt%) target. Under a chamber pressure of 0.3 Pa (Ar/$O_2$ flow: 20/5 sccm), a deposition rate of approximately 6.06 nm/min was achieved, ensuring void-free filling of the 450-nm-thick trenches (widths: 1–3 μm). Although the reflow process resulted in a sidewall tilt angle of approximately 110°, it did not degrade optical mode confinement; conversely, the significantly improved sidewall smoothness effectively suppressed scattering losses. Upon 980 nm excitation, the Er:$Ta_2O_5$ film exhibited intense photoluminescence (PL) centered at 1530 nm (Fig. 1**c**). The emission mechanism follows a standard three-level transition: $Er^{3+}$ ions are excited from the ground state $^4I_{15/2}$) to the $^4I_{11/2}$ state, followed by rapid non-radiative relaxation to the $^4I_{13/2}$ metastable state. The subsequent radiative transition $^4I_{13/2} \rightarrow {}^4I_{15/2}$) yields spontaneous emission spanning 1500–1577 nm. Time-resolved PL measurements (Fig. 1**d**) revealed a fluorescence lifetime of 1.89 ms. Furthermore, the Er:$Ta_2O_5$ waveguide demonstrated a high intrinsic quality (Q) factor of $5.03 \times 10^5$ near 1530 nm, corresponding to a propagation loss of 0.73 dB/cm and a narrow 3-dB resonance bandwidth of 8.4 pm

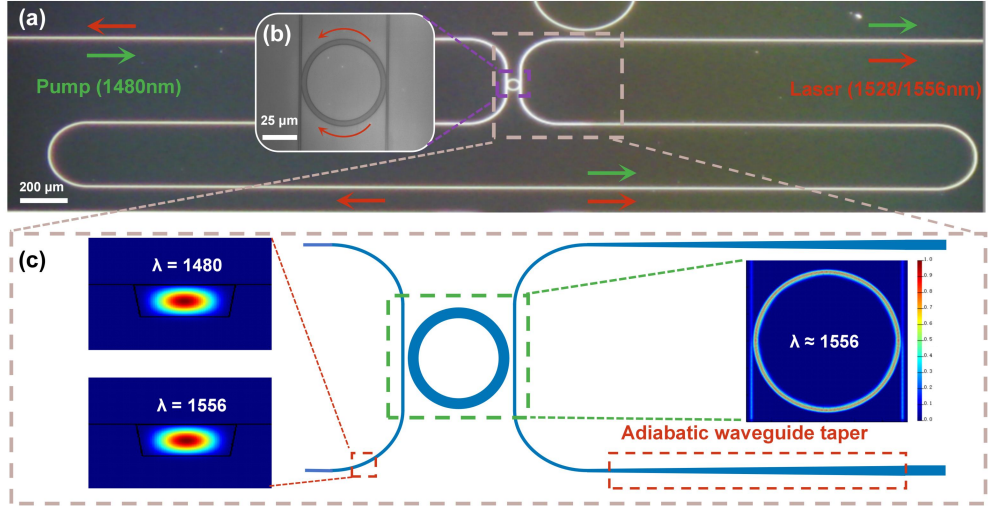

**Fig. 2 a** Optical Microscope Image of Erbium-Doped Tantalum Oxide Laser. **b** Magnified View of the Add-Drop Ring Resonator Mirror; **a** and **b** illustrate the optical paths of the pump and excitation lights. **c** Simplified Schematic Diagram of the Microring Resonant Region and Simulated Diagram of the Optical Mode.

Fig. 2**a** presents a top-view micrograph of the on-chip Er:$Ta_2O_5$ microring laser. The device, occupying a footprint of $6.2 \times 2.3$ mm$^2$, consists of a microring resonator integrated with a U-shaped waveguide. As illustrated in Fig. 2**b**, the waveguide is coupled to the microring at two symmetric points; here, the microring functions as a recirculating element that forms a hybrid resonant cavity with the U-shaped waveguide, utilizing the Er:$Ta_2O_5$ gain medium for optical amplification.To achieve robust laser oscillation within this hybrid architecture, three fundamental conditions must be satisfied:Vernier Spectral Filtering: The microring and the U-shaped waveguide function as two coupled cavities with distinct free spectral ranges (FSRs). The periodic staggering and overlapping of these FSRs induce envelope modulation via the Vernier effect, which is critical for longitudinal mode selection.Phase Matching: The cumulative phase shift over a round trip in the dual-cavity system must be an integer multiple of 2π to ensure constructive interference and resonance.Gain-Loss Balance: Within the $Er^{3+}$ gain bandwidth, the collective gain must compensate for the total round-trip losses, requiring

precise matching between the coupling and loss coefficients.A pump wavelength of 1480 nm was selected to minimize on-chip coupling losses and maximize the mode overlap between the pump and signal fields. Upon injection, the $Er^{3+}$ ions in the U-shaped waveguide absorb pump photons, reaching population inversion and initiating spontaneous emission. As the light circulates within the hybrid cavity, the Vernier effect filters and amplifies only those wavelengths that satisfy the dual-resonance conditions. As the pump power exceeds the lasing threshold, stimulated emission becomes dominant. Subsequent mode competition suppresses auxiliary modes, eventually yielding a stable single-mode laser output.The structural details of the microring resonator are shown in Fig. 2**c**. The gain-region waveguide (3 μm wide, 450 nm thick) utilizes an adiabatic taper to transition to a 1 μm width, effectively stripping higher-order modes to ensure strictly fundamental TE mode operation. Finite-difference eigenmode (FDE) simulations confirm that the effective mode overlap between the 1480 nm pump and the 1556 nm laser signal exceeds 90% at the waveguide bends.Notably, this design employs a non-resonant pumping scheme. While the lasing signal (1500–1577 nm) undergoes strong directional resonant amplification, the 1480 nm pump light remains non-resonant due to the deliberate mismatch in structural parameters at that wavelength. This decoupling ensures that the system is insensitive to the spectral purity or broadening of the pump source, thereby enhancing the stability of the energy conversion process and ensuring efficient single-mode lasing.

Theoretically, the optical field evolution within the hybrid cavity—comprising the U-shaped waveguide and the microring—can be rigorously modeled using the Transfer Matrix Method (TMM). By cascading the transfer matrices of the individual coupling regions and the linear propagation segments, the total cavity response is determined. Specifically, the intermediate optical field amplitude at the projection port, after a single complete round trip, is derived as follows:

$$E'_{through} = E_0 \cdot A_U \cdot e^{i\theta U} \cdot \frac{-k_1^* \cdot k_2 \cdot A_{R/2} \cdot e^{i\theta_{R/2}}}{1 - t_1^* \cdot t_2^* \cdot A_R \cdot e^{i\theta_R}}$$

where represents the initial amplitude, A denotes the loss factor of the waveguide($A = e^{-\alpha L}$, with α being the linear loss coefficient and L the cavity length), K is the propagation constant( $K = \frac{2\pi \cdot n_{eff}}{\lambda}$ ), the phase accumulation of the optical field along the cavity length is $\theta$ ($\theta = KL$), and the complex conjugate terms ($t_1^*, t_2^*, k_2^*$) of the input coupler (transmission coefficient $t_1$, coupling coefficient $k_1$) and output coupler (transmission coefficient $t_2$, coupling coefficient $k_2$) are employed to describe the phase compensation of the reverse optical field.Considering the multiple cyclic superposition of the optical field, the total optical field amplitude at the projection port is derived by summing the geometric series:

$$T = \left| \frac{E_0}{1 - \frac{E'_{through}}{E_0}} \right|^2$$

From this, the transmission spectrum of the hybrid cavity is calculated (as shown in Fig. 4**b**). A hybrid free spectral range (FSR) of approximately 7.12 nm is achieved within the 1520–1580 nm wavelength range, which demonstrates the ability of the Vernier effect to regulate the frequency selection and spacing of resonant peaks .

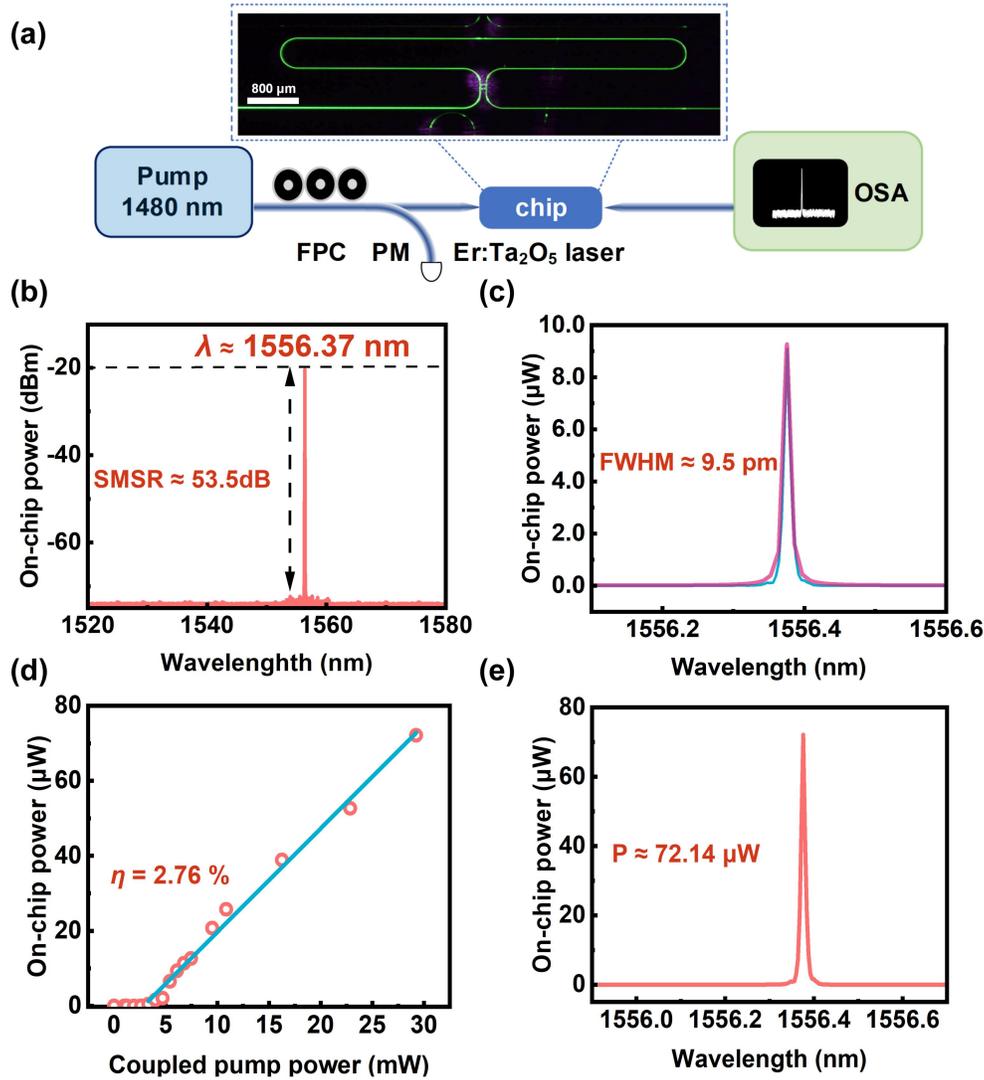

**Fig. 3 a** Experimental Setup Diagram for Testing Laser Performance. **b** Output Spectrum of Erbium-Doped Tantalum Oxide Laser from 1520 nm to 1580 nm. **c** Spectral Magnification and Laser Power Density around 1556 nm; Laser Peaks Fittedwith Lorentzian Line (Blue). **d** Relationship between On-Chip Power of Erbium-Doped Tantalum Oxide Laser and Absorbed Pump Power. **e** Spectral power density of the Er:Ta$_2$O$_5$ microring resonator laser emission.

The experimental configuration for laser characterization is depicted in Fig. 3**a**. Measurements were performed at room temperature without active thermal stabilization. A 1480 nm pump light from a single-mode fiber-coupled laser (SPL-1480-800-FA-B, PatentGuru) was launched into the device via lensed fiber coupling, with the output signal collected using a similar lensed fiber at the opposite facet. The estimated coupling losses were 6 dB/facet at 1550 nm and 7.5 dB/facet at 1480 nm. The emission spectra were characterized using an optical spectrum analyzer (OSA, AQ6375D, Yokogawa Inc.). The inset of Fig. 3**a** captures the visible upconversion fluorescence within the Er:Ta$_2$O$_5$ hybrid cavity under 1480 nm excitation.As shown in Fig. 3**b**, the hybrid cavity's Vernier effect facilitates robust mode selection, yielding a single-mode emission peak at 1556.27 nm across a broad span from 1520 to 1580 nm. The achieved side-mode suppression ratio (SMSR) is as high as 53.3 dB. A high-resolution scan (Fig. 3c) of the 1556 nm peak, fitted with a Lorentzian profile, reveals a full width at half maximum (FWHM) of 9.5 pm. It should be emphasized that this measured linewidth is constrained by the OSA's resolution (0.01 nm), suggesting that the intrinsic physical linewidth is likely significantly narrower.The laser's power characteristics are summarized in Fig. 3d and 3e. Fig. 3d illustrates the on-chip output power as a function of the coupled on-chip pump power. Linear regression indicates a lasing threshold of approximately 3.3 mW and a slope efficiency of 2.76%—the highest value reported to date for Er:$Ta_2O_5$-based lasers. At a pump power of 29.24 mW, the single-ended on-chip output power reaches 72.14 μW. Notably, the total output power could be further enhanced by integrating reflective components, such as a Sagnac loop at the input port, or by extending the gain-region waveguide. Additionally, fiber coupling losses and inherent system losses result in the power collected by the OSA being lower than the actual on-chip power, which must be distinguished in power analysis

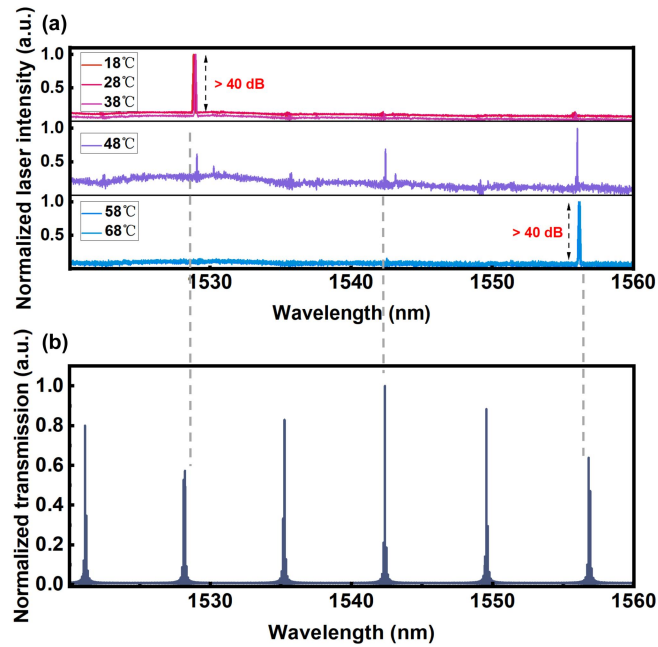

**Fig. 4 a** Temperature-Dependent Diagram of the Laser Obtained via TEC Control. **b** Fitting Diagram of Laser Transmittance versus Wavelength Based on the Design.

Fig. 4a displays the normalized emission spectra of the packaged laser across a temperature range of 18–68 °C (10 °C increments). The pump power was maintained at 30 mW, with precise temperature control achieved via a thermoelectric cooler (TEC). As the temperature rose from 18 °C to 68 °C, the lasing wavelength exhibited a systematic redshift from ~1530 nm to ~1555 nm. This tuning behavior is primarily driven by the synergy between the thermo-optic effect and thermal expansion within the hybrid microring cavity, which collectively modify the resonant condition.Temperature variations also significantly influenced the internal mode competition dynamics. Stable single-mode operation with a side-mode suppression ratio (SMSR) exceeding 40 dB was maintained at most temperatures, with measured wavelengths of 1528.84 nm, 1528.91 nm, and 1529 nm for 18 °C, 28 °C, and 38 °C, respectively, and 1556.18 nm and 1556.2 nm for 58 °C and 68 °C. Notably, at 48 °C, the spectrum exhibited multiple peaks (1529.09 nm, 1544.44 nm, and 1556.03 nm), indicating a transition regime where intensified mode competition allowed multiple resonant modes to simultaneously reach the threshold.To further elucidate these results, Fig 4b presents the simulated passive transmission characteristics of the hybrid cavity derived via the transfer matrix method. The shift trend of the experimental lasing peaks demonstrates high consistency with that of the theoretical transmission peaks, confirming the universal regulatory effect of temperature on the underlying properties of the microring hybrid resonant cavity.While the theoretical peaks serve as a reliable benchmark for predicting lasing wavelengths, slight discrepancies between experimental and simulated positions were observed. These deviations are attributed to the complex interplay of localized gain-loss imbalances and nonlinear thermal gradients not fully captured by the static model.

In summary, we have successfully demonstrated a fully packaged on-chip microring laser on a silicon platform, enabled by a Damascene process tailored for the erbium-doped tantalum oxide Er:$Ta_2O_5$ photonic platform. The Er:$Ta_2O_5$ gain medium exhibited an impressively low propagation loss of 0.73 dB/cm and a high intrinsic quality factor Q of $5.03 \times 10^5$. The device achieved superior performance metrics, including a side-mode suppression ratio (SMSR) of 53.3 dB and a narrow emission full width at half maximum (FWHM) of 9.5 pm. Notably, the laser characterized by a threshold of ~3.3 mW and a slope efficiency of 2.76%—the highest value reported to date for Er:$Ta_2O_5$ lasers. The maximum on-chip output power at a single facet reached 72.14 μW.Future research will focus on further scaling the output power by optimizing fiber-to-chip coupling efficiency, extending the gain-region length, integrating high-reflectivity components, and refining the waveguide geometry to further narrow the intrinsic linewidth. We anticipate that these findings will accelerate the miniaturization and performance enhancement of erbium-based integrated lasers. Furthermore, this work provides a scalable pathway for high-power, single-mode tunable waveguide lasers and establishes a robust technical foundation for the monolithic integration of active and passive photonic structures on the tantalum oxide platform.

**Acknowledgments:** This project was primarily supported by the Key Research and Development Program (2022YFA1404800), National Natural Science Foundation of China (12374359).The authors also thank the Analytical & Testing Center of NPU for providing facilities in EBL, ICP, AFM and SEM measurements.

**Disclosures:** The authors declare no conflicts of interest.

**Data availability:** Data underlying the results presented in this paper are not publicly available at this time but may be obtained from the authors upon reasonable request.